\def\year{2021}\relax
\documentclass[letterpaper]{article} % DO NOT CHANGE THIS
\usepackage{aaai21}  % DO NOT CHANGE THIS
\usepackage{times}  % DO NOT CHANGE THIS
\usepackage{helvet} % DO NOT CHANGE THIS
\usepackage{courier}  % DO NOT CHANGE THIS
\usepackage[hyphens]{url}  % DO NOT CHANGE THIS
\usepackage{graphicx} % DO NOT CHANGE THIS
\usepackage{natbib}  % DO NOT CHANGE THIS AND DO NOT ADD ANY OPTIONS TO IT
\usepackage{caption} % DO NOT CHANGE THIS AND DO NOT ADD ANY OPTIONS TO IT
\usepackage[switch]{lineno}
\usepackage{xcolor}

\renewcommand{\textit}[1]{\emph{#1}}
\newcommand{\posscitet}[1]{\citeauthor{#1}'s (\citeyear{#1})}

\title{Learning Compositional Negation in Populations of Roth-Erev and Neural Agents}
\author {
        Graham Todd\textsuperscript{\rm 1},
        Shane Steinert-Threlkeld\textsuperscript{\rm 2},
        Christopher Potts\textsuperscript{\rm 1} \\
}
\affiliations {
    \textsuperscript{\rm 1} Stanford University \\
    \textsuperscript{\rm 2} University of Washington \\
    gdrtodd@stanford.edu, shanest@uw.edu, cgpotts@stanford.edu
}

\begin{document}
% \linenumbers
\maketitle

\begin{abstract}
Agent-based models and signalling games are useful tools with which to study the emergence of linguistic communication in a tractable setting. 
These techniques have been used to study the compositional property of natural languages, but have been limited in how closely they model real communicators.
In this work, we present a novel variant of the classic signalling game that explores the learnability of simple compositional rules concerning negation. The approach builds on \citet{steinert2016compositional} by allowing agents to determine the identity of the ``function word" representing negation while simultaneously learning to assign meanings to atomic symbols.
We extend the analysis with the introduction of a population of concurrently communicating agents, and explore how the complications brought about by a larger population size affect the type and stability of the signalling systems learned. We also relax assumptions of the parametric form of the learning agents and examine how neural network-based agents optimized through reinforcement learning behave under various task settings.
We find that basic compositional properties are robustly learnable across a wide range of model relaxations and agent instantiations.
\end{abstract}

\section{Introduction}

A symbolic system is \emph{compositional} if the meanings of its complex expressions are determined by the meanings of their constituent parts and how those parts interact. Natural languages are widely assumed to display at least a high degree of compositionality in this sense, and this is often seen as explaining their systematicity and expressivity \citep{Partee84,Janssen97}. At the same time, however, a compositional system requires a degree of abstraction that might prove needlessly burdensome in environments where a simpler scheme would meet the agents' communicative goals.

The question thus arises: under what conditions can compositional systems emerge? Building on work by \citet{skyrms2009evolution}, \citet{steinert2016compositional} addresses this question using extensions of the \emph{signalling systems} of \citet{Lewis69}, which model how arbitrary symbols can acquire meaning through simple, reward-driven interactions between abstract senders and receivers. \citeauthor{steinert2016compositional} defined the \emph{Negation Game}, a variant of the classic signalling game that allowed agents to make use of simple compositional rules concerning negation, and showed that under certain settings it was advantageous to do so. 

The present paper revisits \citeauthor{steinert2016compositional}'s Negation Game, seeking to increase the realism of the compositional rules studied therein, and to more fully characterize the conditions under which compositionality can arise. In particular, we focus on two restrictions that \citeauthor{steinert2016compositional}'s Negation Game imposes. First, in the Negation Game, agents innately agree to reserve a particular symbol for communicating negation, which simplifies the learning task considerably. Second, the Negation Game (like most signalling game variants) involves just two agents, whereas coordinating convention formation across multiple disparate actors is both more realistic and more challenging. To explore the extent to which emergent compositonality depends on these restrictions, we propose a new \emph{Learned Negation Game} in which populations of agents communicate without prior knowledge of the identity of the negation symbol. Our central finding is that compositionality emerges even in this much less restricted setting.

We further extend our analysis by relaxing assumptions about the parametric form of the learning agents. Our first experiments, and many prior studies, make use of simple Roth-Erev learning agents \citep{roth1995learning}. While effective in many contexts, Roth-Erev agents encode a substantial amount of prior knowledge about the environment and learning targets. Is this necessary for emergent compositionality? To address this question, we examine the feasibility of learning compositional rules for populations of neural network-based agents trained with a reinforcement learning objective. Strikingly, even these agents are capable of learning and utilizing simple compositional rules. 

Finally, we introduce the  \emph{Combined Negation Game}, which merges the Learned Negation Game with \citeauthor{steinert2016compositional}'s Functional Negation Game. In the Combined Negation Game, agents in the population must learn not only the identity of the negation symbol but also its meaning. This results in a substantially more challenging environment that presumably comes even closer to the conditions under which compositionality is acquired in natural languages. While the increased difficulty of this task does result in lower performance across both types of learners, we still observe the acquisition of simple compositionality in a variety of contexts.

Through our experiments, we find that compositional negation is robust to a large range of model relaxations and agent instantiations. This, in turn, bolsters claims that the emergence of compositional signalling is not a mere artifact of specific model implementations, but rather a consistently useful strategy when communicating in complex environments.

\section{Atomic Signalling Games}

The classic (or ``atomic'') signalling game described by \citet{Lewis69} consists of two agents, a \textit{sender} and a \textit{receiver}.
At each play of the game, a \textit{state} $s$ is selected from $\{s_1, s_2, \ldots, s_n\}$. The state is passed to the sender, which selects a \textit{symbol} $m$ from $\{m_1, m_2, \ldots, m_k\}$. The receiver is then given the symbol (but not the state) and selects an \textit{action} $a$ from $\{a_1, a_2, \ldots, a_n\}$. If the action matches the state (that is, if $s = s_i$ and $a = a_i$ for some $i \in \{1, 2, \ldots, n\}$), the communication is said to be successful and each agent receives a joint reward. If the action differs from the state, the communication is unsuccessful and no reward is given.

Crucially, for the atomic signalling game, none of the symbols have any \textit{a priori} meaning. Rather, their meanings are constructed through the joint action of the sender and receiver. An optimal communication scheme is simply an arbitrary assignment of symbols to states. As long as each state is given a unique symbol, the pair of agents will be able to achieve a perfect rate of communication success. Over time, even quite simple learning mechanisms are capable of guiding the sender and receiver agents to an efficient communication scheme.

While powerful, the atomic signalling game seems to leave out many elements of emergent communication we might consider important. For instance, it does not account for the complicating force of population dynamics. Real communicators do not need to coordinate only with a single other agent and in a single role, but rather with a population of agents, and in both sender and receiver roles. Introducing even a single other agent to the Lewis signalling game already represents a substantial increase in difficulty. With a population of agents, it is no longer always sufficient for an agent to greedily maximize its probability of communication success with its current partner, as doing so might cause it to have a lower rate of communication success with the rest of the population, should they be converging to a different communication scheme.

\section{\posscitet{steinert2016compositional} Negation Games}

Another important aspect of communication left out of the atomic signalling game is compositionality. Because agents send and receive only a single symbol, there is no chance to study the dynamics that arise from the composition of multiple symbols. There are numerous ways to remedy this by modifying the atomic signalling game, but of particular interest to us are the \textit{Basic} and \textit{Functional} Negation Games proposed by \citet{steinert2016compositional}, which probe the compositionality required for a logical negation operator. 

Like the atomic signalling game, the Basic $n$-Negation Game comprises a single sender and receiver. The Negation Game seeks to capture the intuition that states might have natural opposites or negations (for instance, the state corresponding to the concept of ``danger'' might be opposite to the state corresponding to the concept of ``safety''), and that agents who take advantage of this structure when communicating are likely to be more  successful than agents who don't. A derangement $f: [2n] \rightarrow [2n]$ (a bijection with no fixed points) is used to map from $n$ states to their negations, resulting in a total of of $2n$ states. Unlike the atomic signalling game, however, agents in the Negation Game do not have access to as many symbols as states in the environment. Rather, they communicate with $n$ normal symbols and one special ``negation symbol'', for a total of $n+1$ symbols.

The game plays out as follows: a state $s$ is selected from $\{s_1, s_2, \ldots, s_{2n}\}$. The sender uses the state to select a symbol from $\{m_1, m_2, \ldots, m_n\} + \{ \neg \}$, where $\neg$ represents the special negation symbol. If the symbol selected is \textit{not} the negation symbol, then the symbol is passed to the receiver, which selects an action as in the atomic signalling game. If the symbol selected \textit{is} the negation symbol, then the sender selects an additional non-negation symbol $m$ from the distribution corresponding the negation of the current state, or $f(s)$. The message sent to the receiver is then $\neg m$. In this case, the receiver first selects an action $a$ from $\{a_1, a_2, \ldots, a_{2n}\}$ corresponding to the distribution obtained by processing symbol $m$, but then actually \textit{performs} the action $f^{-1}(a)$. If the performed action matches the original state, then both agents receive a reward.

In the both variants of the Negation Game, perfect communication success requires agents to make use of the negation symbol in order to be able to communicate about all states. Failure to use the negation symbol would result in, at best, a communication success rate of 0.5. \citet{steinert2016compositional} finds that even simple learning agents are capable of achieving rates of communication success between 0.851 and 0.605 for values of $n$ between $2$ and $8$ after $10,000$ iterations of Roth-Erev learning.

While the Negation Game is a powerful tool for examining how agents can acquire a basic compositional ability, it is not without its shortcomings. We note that it encodes the following two assumptions that are worth further interrogation:

\begin{enumerate}
    \item Agents inherently agree about the \textbf{identity} of the special negation symbol.
    \item Agents inherently agree about the \textbf{function} of the special negation symbol.
\end{enumerate}

\citet{steinert2016compositional} seeks to address the second assumption with the functional variant of the Negation Game, in which agents must learn to define the negation symbol by selecting its meaning from a short list of possible functions, in addition to determining when to deploy it. We revisit these ideas in our Combined Negation Game. Before this, though, we seek to address the first assumption with our Learned Negation Game.

\section{The Learned Negation Game}

Our Learned Negation Game adds a relatively straightforward complication to the standard Negation Game: instead of the negation symbol having a reserved identity that is known and shared by each agent before the game, agents instead assign the role of negation to an arbitrary symbol and must learn to agree on the negation symbol's identity. Like the Negation Game, the Learned Negation Game makes use of a derangement function $f$ to map from $2n$ states to their negations. Similarly, there are $n+1$ possible symbols, though there is no distinction drawn between ``normal'' symbols and the negation symbol. 

The game plays out as follows. First, the sender selects its negation symbol (the symbol it believes to represent minimal negation), which we call $\phi$, from $\{m_1, m_2, \ldots, m_{n+1}\}$ and the receiver does the same for its negation symbol, which we call $\psi$. A state $s$ is selected from $\{s_1, s_2, \ldots, s_{2n}\}$. Using the state, the sender selects a symbol from $\{m_1, m_2, \ldots, m_{n+1}\} - \{\phi\} \cup \{\neg\}$. Here, $\neg$ no longer serves as the actual negation symbol, but rather as the ``abstract negation'' symbol. That is, the symbol $\neg$ is never actually sent to the receiver. Instead, it acts as a stand-in for whatever symbol was currently selected by the sender to act as the negation symbol. So, if the sender selects $\neg$ as its first symbol, a second symbol $m$ is drawn from the distribution corresponding to $f(s)$, but the message sent to the receiver is actually $(\phi, m)$. (This allows for the sender to reliably ``point to" and send the negation symbol, regardless of which identity was selected for it in a given episode.)

In this case, the receiver has three options, depending on the value of $\psi$. If $\psi = \phi$ (that is, the receiver believes the same symbol to act as negation as the sender), then the receiver draws an action $a$ from the distribution corresponding to $m$ and performs the action $f^{-1}(a)$. Conversely, if $\psi = m$, then the receiver draws $a$ from the distribution corresponding to $\phi$ and similarly performs $f^{-1}(a)$. We note that this means the order of symbols does not matter -- from the receiver's perspective, sometimes the negation symbol precedes the other symbol, and sometimes it follows. In some instances, however, $\psi \not = \phi$ and $\psi \not = m$. Given the explicit grammar of the Basic Negation and Learned Negation Game (messages are of fixed length 1, unless the negation symbol is being used), this represents an unparseable message for the receiver. In such cases, the receiver randomly selects either $\phi$ or $m$ and performs an action $a$ drawn from the distribution corresponding to the selected symbol.

In the case where the first symbol selected is not $\neg$, then the message $(m)$ is passed to receiver, which simply draws and performs an action $a$ from the distribution corresponding to $m$.

The Learned Negation Game intuitively models the case in which agents are innately capable of performing the minimal negation operation (as represented by the use of the derangement function) but must learn how to assign this operation to a symbol. We can consider this the converse of the Functional Negation Game, which models the case in which agents innately share knowledge of \textit{which} symbol denotes the relevant function word, but the meaning of that function must be learned. 

We are interested in interrogating the dynamics of the Learned Negation game played not just between a single sender and receiver, but between populations of agents capable of performing both roles. For the following experiments, we examine the behavior of two different kinds of learning algorithms on the population-based Learned Negation game.

\section{Roth-Erev Learners}

Our first set of simulations focus on Roth-Erev agents \citep{roth1995learning}. We first define these agents for our Learned Negation Game, and then we describe our simulations.

\subsection{Agent Definition}

For the atomic signalling game, a Roth-Erev learner comprises two matrices of integer counts: a \textit{sender matrix} and a \textit{receiver matrix}. Each matrix tracks the \emph{accumulated reward} for that agent accrued by selecting a given symbol for a given state (in the case of the sender matrix) or by selecting a given action for a given symbol (in the case of the receiver matrix). The selection of symbols is governed by the distribution of accrued rewards. For instance, when an agent in the sender role is given a state $s$, it consults the row of its sender matrix corresponding the state, normalizes the vector of accrued rewards for each symbol, and draws a symbol from the resulting distribution. A mirrored operation occurs for an agent in the receiver role when it is given a symbol $m$, as it normalizes the accrued rewards for each possible action and samples from the resulting distribution. Learning in the Roth-Erev setting is handled with incrementation: when an agent successfully communicates, it adds $\eta$ for the accrued reward of the selected symbol or action given the received state or symbol, where $\eta$ is the \textit{learning rate}.

While extremely simple, Roth-Erev learning remains powerful enough to solve many instantiations of communication games. Indeed, \citet{steinert2016compositional} demonstrates analytically that Roth-Erev learning is capable of solving both the Negation Game and the Functional Negation Game. However, in order to accommodate the Learned Negation Game, an additional set of parameters must be added, here referred to as the \textit{negation distribution}. The negation distribution tracks the agent's accrued reward for selecting each symbol as the negation symbol. The reward for this selection is shared across the sender and receiver roles, since the decision is made in both. To select the identity of the negation symbol, an agent simply draws from the distribution of accrued rewards.

Finally, in order to improve the performance of Roth-Erev learning, we introduce the notion of \textit{reward resetting}. Intuitively, we occasionally reduce the accumulated rewards of the agents while keeping their relative proportions the same (details on implementation available in the Technical Appendix). One attribute of Roth-Erev learning is that as accumulated rewards accrue over time, the relative impact of a single training episode becomes less and less. This can lead to situations where agents adopt sub-optimal strategies but are extremely unlikely to find new strategies, since doing so would require repeatedly selecting an action with a lower accrued reward. We find empirically that reward resetting often helps Roth-Erev agents escape from local optima during training. 

\subsection{Experimental Details}
We examine the performance of a population of Roth-Erev agents on the Learned Negation Game for $n = 4$ and $n = 8$ (corresponding to $8$ and $16$ states) and for populations sizes $p = 2$ and $p = 5$. Each experiment consists of $e$ \textit{learning events}. We define a learning event as follows. First, we shuffle the order of agents. Second, each agent plays $k$ games against both itself and all other agents, in both roles. We call this parameter $k$ the number of \textit{learning trials}. We perform reward resetting on each agent in the population every $r$ learning events. We report the \textit{fitness} of the agent population, defined as the average rate of communication success across all pairings of agents in both roles, sampled across $g$ games in which learning is frozen. 

For the following experiments, we set $\eta=1$, $e=10000$, $k=10$, $r=1000$, and $g=50$. Each experiment was also repeated for 10 repetitions, with a random seed of $42+i$ applied to each, with $i= 0, 1, \ldots, 9$. In all experiments, fitness evaluations were performed every 100 learning events.

Under our definitions, even a population size of 2 is not equivalent the single sender, single receiver setting common in the literature. This is due to the fact that each agent performs both the sender and receiver roles. Additionally, since some parameters are shared between an agent in the sender and receiver roles (namely, the negation symbol distribution), the fitness might be artificially inflated by an agent's increased performance in playing against itself. In practice, we do not see large differences between these settings, so we concentrate on fitness with self-play and present the results without self-play in the Technical Appendix.

\begin{table}[tp]
    \centering
    \begin{tabular}{|l|l|l|}
    \hline
        & $n=4$     & $n=8$     \\ \hline
    $p=2$ & 0.931 {\small (0.897, 0.964) \par} & 0.877 {\small (0.842, 0.911) \par} \\ \hline
    $p=5$ & 0.935 {\small (0.884, 0.987) \par} & 0.835 {\small (0.797, 0.873) \par} \\ \hline
    \end{tabular}
    \caption{Peak fitness for populations of Roth-Erev agents in the Learned Negation game, averaged over 10 repetitions. 95\% confidence intervals are shown in parentheses. For comparison, with $n=4$ and $n=8$, a random policy leads to a fitness of $0.125$ and $0.0625$, respectively. Failure to utilize minimal negation would result in a maximum possible fitness of $0.5$.}
    \label{avg-roth-erev-learned-neg}   
\end{table}

\subsection{Results}
\begin{figure}[tp]
    \centering
    \includegraphics[width=1\linewidth]{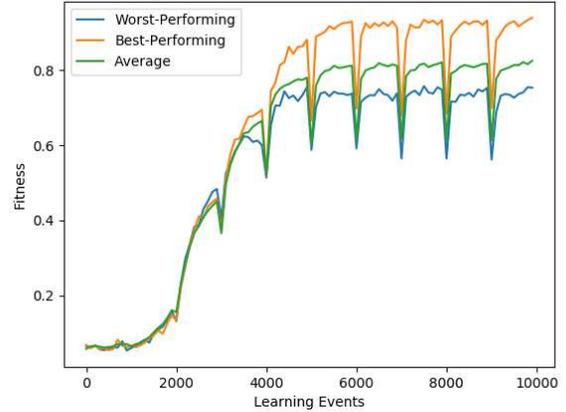}
    \caption{Fitness over time in the Learned Negation Game for a population of 5 Roth-Erev agents with $n=8$. Performance is shown for the individual best- and worst- performing repetitions as well as the average over all 10 repetitions. The trajectory is characterized by a rapid initial increase followed by plateauing and periodic dips, which are caused by the \textit{reward resetting}. Notably, fitness rapidly returns to its previous levels following each reset.}
    \label{fig:roth-erev-plot}
\end{figure}

Table~\ref{avg-roth-erev-learned-neg} presents the peak fitness (with self-play) of the Roth-Erev agents for specified values of $n$ and $p$, averaged across 10 repetitions, and with 95\% confidence intervals. We choose to show the peak fitness of the agents (as opposed to their final fitness) because we are primarily interested in demonstrating that Roth-Erev agents have the \textit{capacity} to solve the learned-negation signalling game. That being said, the fitness trajectory of the population of agents is still of some interest. In Figure~\ref{fig:roth-erev-plot}, we show the population fitness (including self-play) over time for the best- and worst- performing repetitions of the experiment with $n=8$ and $p=5$, as well as the average fitness across all repetitions. 

In all the settings above, the population of Roth-Erev agents is capable of achieving high rates of communication success. In particular, a fitness significantly larger than 0.5 indicates that the agents have learned to make use of the minimal negation operator. Fitness is decreased for larger values of $n$, as expected. However, more surprisingly, we find that increasing the population size from $2$ to $5$ does not significantly decrease the average peak performance. This offers some support for our hypothesis that larger population sizes, while a challenge for learning agents, are not insurmountable obstacles. 

The fitness plot in Figure~\ref{fig:roth-erev-plot} shows a fairly steady increase in performance over the first 5000 learning events, before most runs reach a plateau. The periodic dips in performance are caused from the reward resetting, though it is notable that fitness always quickly returns to its pre-reset levels. Also of note is the relatively slow learning that takes place here, especially when compared to the results presented in \citet{steinert2016compositional} on the Basic Negation Game. Given the similarity of the models in question, it seems likely that this can be attributed to the increased difficulty of Learned Negation Game compared and the challenges posed by populations of agents interacting.

While the results presented here of course only paint out a small slice of the overall parameter space for the learned negation game, we believe they present reasonable evidence that, despite contending with relaxed assumptions in the signalling game, even relatively simple agents are still capable of learning basic compositional rules concerning negation.

\section{Neural Network-Based Learners}

While we have demonstrated that Roth-Erev learning is often sufficient for populations of agents to converge on the Learned Negation game, it is not without its shortcomings. Of note, Roth-Erev learning agents explicitly encode which parameters are to be used for each given state or symbol and, hence, information gained in one training episode is of no use in other episodes in which the received state or symbol is different. We might expect that real communicators are better able to synthesize experience across a variety of situations, and that their understanding of the problem space is represented across a variety of parameters. To that end, we also investigate the behavior of senders and receivers that have been instantiated with an artificial neural network and trained using reinforcement-learning algorithms.

\subsection{Agent Definition}

Our neural-network learner in the atomic signalling game comprises six components: a \textit{sender embedding}, a \textit{receiver embedding}, a shared \textit{processor MLP}, a \textit{sender projection}, a \textit{receiver projection}, and a shared \textit{critic MLP}.

An agent's behavior in the sender role is as follows: the given state is fed into the sender embedding in order to get a hidden representation of the state. This hidden representation is then fed through the \textit{processor}, which is a multi-layer perceptron with layer normalization \cite{ba2016layer} and a \texttt{tanh} non-linearity. This modifies the hidden representation, which is then passed through the \textit{sender projection}, which outputs a score for each possible symbol. The scores are passed through a \texttt{softmax} function, and a symbol is drawn from the resulting distribution. At the same time, the sender embedding (before the processor) is passed through the \textit{critic} (a multi-layer perceptron with the same shape and features as the processor) to produce a value estimate of the current state. 

When in the receiver role, an agent behaves in much the same way, except that the receiver embedding and projection layers are used to process an incoming symbol, and an action is ultimately selected.

Unlike in the Roth-Erev case, agents share parameters across their sender and receiver roles. In particular, the processor and the critic systems are shared. This captures the intuition that an effective communicator ought to be able to leverage its experience as both a sender and receiver in order to help it perform both roles.

As with the Roth-Erev case, an additional component is necessary for the neural-network learner to be able to complete the Learned Negation Game: a \textit{negation distribution MLP} is added to the model alongside a \textit{negation symbol critic}. The negation distribution MLP maps from a static input to a distribution over symbols (excluding the abstract negation symbol). The negation symbol critic is a multi-layer perceptron that maps from a selected negation symbol to the expected value of that selection.

In all variants of the game, the neural-network learners are optimized using Proximal Policy Optimization \cite{schulman2017proximal}, an advantage-based actor-critic reinforcement learning system. Each agents tracks the decisions made and rewards accrued for each role separately, though once again choices about the negation symbol in the Learned Negation Game are made in both roles. 

\subsection{Experimental Details}
As with the Roth-Erev case, we examine the fitness of a population of neural agents for $n \in \{4, 8\}$ and $p \in \{2, 5\}$. We use the same definitions of learning events and fitness, and similarly set $e = 10000$, $k = 10$, $r = 1000$, $g=50$, and random seeds of $42, 43, 44, \ldots, 51$.

For each of the neural agents, the embedding dimension for both the sender and receiver systems was 128. The \textit{processor}, \textit{critic}, \textit{negation distribution}, and \textit{negation critic} systems consisted of 5 layers of dimension 128. The learning rate was set to $0.002$, and the \texttt{Adam} optimizer was used with $\textit{betas} = [0.9, 0.999]$. For PPO-specific hyperparameters, we set $\epsilon = 0.2$ and $K = 4$. (Since all episodes consist of only a single interaction, the value of $\gamma$ is irrelevant.)

\subsection{Results}
\begin{figure}
    \centering
    \includegraphics[width=1\linewidth]{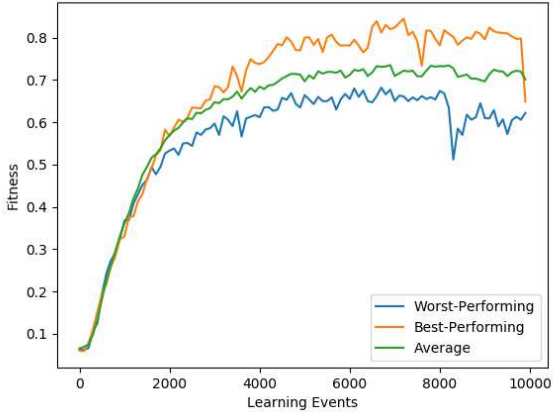}
    \caption{Fitness over time in the Learned Negation Game for a population of 5 neural agents with $n=8$. Performance is shown for the individual best- and worst-performing repetitions, as well as the average over all 10 repetitions. The trajectory is similar to that of the Roth-Erev agents, though in general the neural agents seem slightly less stable.}
    \label{fig:neural-plot}
\end{figure}

Table~\ref{avg-neural-learned-neg} presents the peak fitness (with self-play) of the neural agents for the specified values of $n$ and $p$, averaged across 10 repetitions. Mirroring the Roth-Erev case, we also show the fitness trajectory for the best- and worst- performing repetitions, as well as the average overall, for $n=8$ and $p=5$ in Figure~\ref{fig:neural-plot}.

\begin{table}[tp]
    \begin{center}
    \begin{tabular}{|l|l|l|}
    \hline
        & $n=4$     & $n=8$     \\ \hline
    $p=2$ & 0.849 {\small (0.750, 0.948) \par} & 0.756 {\small (0.683, 0.828) \par} \\ \hline
    $p=5$ & 0.872 {\small (0.816, 0.927) \par} & 0.762 {\small (0.738, 0.801) \par} \\ \hline
    \end{tabular}
    \caption{Peak fitness for populations of neural agents in the Learned Negation game, averaged over 10 repetitions. 95\% confidence intervals are shown in parentheses. For $n=4$ and $n=8$, a random policy leads to a fitness of 0.125 and 0.0625, respectively. Failure to utilize minimal negation would result in a maximum possible fitness of 0.5.}
    \label{avg-neural-learned-neg}
    \end{center}
\end{table}

We broadly observe that neural agents are capable of achieving similarly high performance to the Roth-Erev agents, though in general we observed the average fitness to be slightly lower. However, the hyperparameters for the neural experiments were chosen heuristically and, like those of most deep reinforcement learning models, our results are somewhat fragile. In particular, we occasionally observed that fitness would reach a local optima at around 0.3 and remain there for the duration of the repetition. It seems possible that the population of agents fell into a suboptimal \textit{partial pooling equilibrium}, as discussed in \citet{skyrms2009evolution}. In future work, we plan to more systematically explore these hyperparameters, to better understand their relationship to the emergence of compositionality.

The fitness trajectory in Figure~\ref{fig:neural-plot} tells a similar story to that of the Roth-Erev agents: a relatively fast increase in fitness during the first half of the experiment, followed by a gradual plateauing. We further note that the neural agents seem slightly less stable than the Roth-Erev agents, though any potential effect there seems minor.

In sum, we find that, despite encoding less information about problem structure than Roth-Erev learners, populations of neural network-based agents are still successful in the Learned Negation Game.

\section{The Combined Negation Game}
As previously mentioned, the Learned Negation Game addresses the assumption of the Negation Game that agents inherently agree on the identity of the special negation symbol, while \posscitet{steinert2016compositional} Functional Negation game addresses the assumption that they inherently agree on its meaning. It is also possible to merge these two variants of the negation game to further weaken the implicit assumptions of the model. We call this novel variant the Combined Negation Game. The present section begins to explore this game experimentally using both Roth-Erev learners and neural network learners.

\subsection{Game Definition}

The Functional Negation Game is described in detail in \citet{steinert2016compositional}. In brief, the idea is to allow agents the flexibility to interpret the negation symbol $\neg$ as they wish, with the hypothesis that interpreting it as minimal negation is the optimal choice. In our case, the Combined Negation Game differs slightly from the Functional Negation Game described by \citeauthor{steinert2016compositional} and extends the Learned Negation Game as follows. At the start of each episode, in addition to selecting the identity of the negation symbol, the sender and receiver each independently select its meaning from a list of three options: \textbf{ignore}, \textbf{atomic}, and \textbf{negation}. Here, \textbf{ignore} represents the case in which agents completely disregard the negation symbol, \textbf{atomic} the case in which agents interpret the negation symbol as a standard atomic symbol, and \textbf{negation} the case in which agents understand the negation symbol to represent the minimal negation operation. The behavior of each agent depends on its selection for the meaning of the negation symbol.

For the sender, if the meaning selected is either \textbf{ignore} or \textbf{atomic}, then it selects a single symbol as in the atomic signalling game, with no chance to produce a second symbol. Intuitively, this captures the notion that since neither the \textbf{ignore} function nor the \textbf{atomic} function require any sort of compositionality, the sender has no reason to compose two symbols. If the sender selects the meaning of \textbf{negation}, then it plays out exactly as in the Learned Negation Game.

For the receiver, the choice of meaning matters only in the case where it receives two symbols and one of them matches its selected negation symbol. If the receiver selects \textbf{ignore}, then it simply interprets the other symbol of the message without applying the inverse derangement function (i.e., as in the atomic signalling game). If the receiver selects \textbf{atomic} as its meaning, then it interprets the negation symbol itself without applying the inverse derangement, ignoring the other symbol in the message. Only in the case where the receiver selects \textbf{negation} does it interpret the other symbol of the message using the inverse derangement function, as in the Learned Negation Game.

The Combined Negation Game adds one more decision that the communicating agents must cohere on in order to succeed. That is, in order to be able to communicate about all states in the environment, they must agree to treat the negation symbol as the minimal negation operation, agree on which symbol is to be used as the negation symbol, and agree on the mapping from states to symbols for the remainder of the vocabulary. Because of this, the combined negation encodes only minimal assumptions about the abilities and prior knowledge of the communicators. 

\subsection{Experimental Details}

In order to implement the Combined Negation Game for Roth-Erev learners, we must simply add an additional vector of accumulated rewards, one for each of the three possible meanings of the negation symbol. In order to implement the Combined Negation Game for neural network-based learners, we add a \textit{function distribution MLP} and a \textit{function meaning critic}, which map respectively from a static input to a distribution over function meanings and from the selected meaning to its expected value. The architectures of these components are equivalent to the negation distribution MLP and the negation symbol critic, respectively.

The experiments were once again carried out for $n \in \{4, 8\}$ and $p \in \{2, 5\}$, using the exact same settings for all hyperparameters as in our previous simulations.

\subsection{Results}

\begin{table}[tp]
    \centering
    \begin{tabular}{|l|l|l|}
    \hline
        & $n=4$     & $n=8$     \\ \hline
    $p=2$ & 0.844 {\small (0.754, 0.933) \par} & 0.777 {\small (0.681, 0.873) \par} \\ \hline
    $p=5$ & 0.710 {\small (0.622, 0.797) \par} & 0.671 {\small (0.592, 0.750) \par} \\ \hline
    \end{tabular}
    \caption{Peak fitness for populations of Roth-Erev agents in the Combined Negation game, averaged over 10 repetitions. 95\% confidence intervals are shown in parentheses. The rates are lower than for the Learned Negation Game (Table~\ref{avg-roth-erev-learned-neg}), reflecting the more challenging setting.}
    \label{avg-roth-erev-combined-neg}    
\end{table}

\begin{table}[tp]
    \centering
    \begin{tabular}{|l|l|l|}
    \hline
        & $n=4$     & $n=8$     \\ \hline
    $p=2$ & 0.809 {\small (0.733, 0.885) \par} & 0.670 {\small (0.584, 0.755) \par} \\ \hline
    $p=5$ & 0.638 {\small (0.609, 0.668) \par} & 0.567 {\small (0.494, 0.640) \par} \\ \hline
    \end{tabular}
    \caption{Peak fitness for population of neural agents in the Combined Negation game, averaged over 10 repetitions. 95\% confidence intervals are shown in parentheses.}
    \label{avg-neural-combined-neg}    
\end{table}

The results in Table~\ref{avg-roth-erev-combined-neg} and Table~\ref{avg-neural-combined-neg} mirror those in Table~\ref{avg-roth-erev-learned-neg} and Table~\ref{avg-neural-learned-neg}, respectively. Generally, we observe that average fitness on the Combined Negation Game remains relatively high in all settings and for both types of learners, with the notable exception of the neural learners in the $n=8$, $p=5$ setting. It seems likely that the additional axis of coordination required by the Combined Negation Game proved challenging for the slightly less stable learning of the neural agents. Even for $n=8$ and $p=5$, however, the maximal peak fitness achieved by the neural agents across the 10 repetitions was 0.838, indicating that the neural agents are capable of learning the Combined Negation Game in this setting. 

Overall, however, fitness is decreased compared to the Learned Negation game. This is unsurprising, given the further relaxed assumptions of the Combined Negation game. Compared to the Learned Negation game, we also observe a more substantial decrease in fitness going from a population of two agents to a population of five. We have previously hypothesized that larger population sizes increase the difficulty for agents to coordinate their decisions. It seems reasonable, then, thatthe added dimension of coordination in the Combined Negation game (since agents must agree that the minimal negation operator is helpful for communication before they can even being to learn how to use it) further amplifies the deleterious effect of larger population sizes, when compared to the Learned Negation game.

\section{Related Work}

Our work draws most directly on the studies presented in \citet{steinert2016compositional}. \citeauthor{steinert2016compositional} uses the Basic and Functional Negation Games to present an evolutionary argument for why many natural languages seem to have the property of compositionality. We extend the argument presented there by relaxing some of the core assumptions (i.e., prior knowledge of agents, number of agents communicating) baked into the Negation Game, but similarly find the emergence of compositional signalling systems.

Prior work in signalling games has also often concerned multiple agents. \citet{nowak1999evolution} present an evolutionary game-theoretic approach in which an agent's fitness is determined by its ability to communicate and the evolutionary steps of selection, reproduction, and mutation slowly drive the population's fitness up over the course of many generations. Our work differs in that it implicitly models only a single generation of agents. In an evolutionary simulation, low-performing individuals are gradually removed from the population, allowing better strategies to take their place. By contrast, in our work each agent must independently learn to adopt an efficient (i.e., compositional) communication system in order for overall fitness to be high. In practice, both types of simulation provide valuable but distinct insights into the emergence of communication.  

\citet{skyrms2009evolution} presents other types of signalling games with multiple agents. In the simplest setting, a pair of sender agents must cooperate in order to transmit information about the state to a single receiver, as each individually is capable of encoding only part of the state (a similar game model is deployed by \citet{barrett2018hierarchical}). In another game, a single sender emits a message to a pair of receivers that must learn to coordinate. Our work similarly explores the challenges of coordinating between multiple communicators, but does so in a less structured fashion. The population-based games described above makes no distinction between agents in particular roles and additionally impose no requirements about the specific number of agents involved in the communication game. 

Reinforcement learning in signalling games is also well studied  \citep{barrett2006numerical, argiento2009learning}, and prior work has also considered the effect of relaxing assumptions on the parametric form of the communicator agents. \citet{catteeuw2013limits}, for instance, explore the performance of Roth-Erev reinforcement learning, learning automata, and Q-learning on the atomic signalling game and find that all three types are capable of quickly converging to a efficient signalling system. This analysis, however, is restricted to signalling games without compositionality and between only two agents. We extend the work by relaxing these two constraints.

Finally, \citet{mordatch2018emergence} present a grounded model of multi-agent emergent communication using neural network-based reinforcement learners. The model allows agents to both produce linguistic messages and communicate non-linguistically using cues pointing or motion. \citeauthor{mordatch2018emergence} find that agents are capable of learning basic compositional rules (for instance, understanding the meaning ``red square'' using two separate atomic symbols). Our work more specifically examines compositional negation and the effects of various modeling assumptions on the acquisition of such a compositional ability.

\section{Conclusion and Future Work}
We have shown that simple compositional systems can emerge from the interactions of populations of agents in a wide variety environments. We systemically relaxed assumptions encoded in prior agent-based studies of compositional negation and arrived at the novel Learned Negation Game and Combined Negation Game. In studies of both Roth-Erev and neural agents, we found that compositional communication schemes can form, though larger population sizes and relaxed assumptions unsurprisingly do tend to reduce rates of communication success. In future studies, we seek to explore the particular dynamics of the Learned and Combined Negation Games in more detail. We are also interested in exploring the interactions between learning and evolution in compositional signalling games.

\bibliography{bibliography}

\newpage
\clearpage

\section*{Technical Appendix}

\setcounter{table}{0} \renewcommand{\thetable}{\arabic{table}}

\subsection{Reward Resetting}
Resetting the accumulated rewards for the Roth-Erev agents is a fairly straightforward process. We perform Reward Resetting independently on each agent's rewards accrued in the sender and receiver roles, as well as the rewards accrued for selecting the identity of the negation symbol and the meaning of the negation symbol (in the Combined Negation Game).

To perform Reward Resetting normalize the accrued rewards into a distribution. The values of this distribution are then multiplied by a constant (the "initial reward"), which for our experiments was set to 100. We then add a smoothing factor of 1 to each of the resulting values, and use them to replace the existing accumulated rewards.

\subsection{Fitness Values Excluding Self-Play}
Below we present the fitness of the population of agents evaluated without self-play (i.e., when an agent plays either game in both the sender and receiver role simultaneously) for each of $n = 4, 8$ and $p = 2, 5$, for both Roth-Erev and neural agents on both the Learned Negation Game and the Combined Negation Game. Note that self-play is still performed during training in all cases and that the performance of random agents and agents that do not use the negation symbol remains the same as in the main experiments.

\begin{table}[htb]
    \centering
    \begin{tabular}{|l|l|l|}
    \hline
        & $n=4$     & $n=8$     \\ \hline
    $p=2$ & 0.941 {\small (0.910, 0.964) \par} & 0.887 {\small (0.851, 0.923) \par} \\ \hline
    $p=5$ & 0.936 {\small (0.885, 0.987) \par} & 0.836 {\small (0.798, 0.873) \par} \\ \hline
    \end{tabular}
    \caption{Peak fitness for populations of Roth-Erev agents in the Learned Negation game, averaged over 10 repetitions, and excluding self-play. 95\% confidence intervals are shown in parentheses.}
    \label{avg-roth-erev-learned-neg-no-self}   
\end{table}

\begin{table}[htb]
    \centering
    \begin{tabular}{|l|l|l|}
    \hline
        & $n=4$     & $n=8$     \\ \hline
    $p=2$ & 0.861 {\small (0.767, 0.951) \par} & 0.775 {\small (0.707, 0.843) \par} \\ \hline
    $p=5$ & 0.873 {\small (0.817, 0.928) \par} & 0.770 {\small (0.737, 0.803) \par} \\ \hline
    \end{tabular}
    \caption{Peak fitness for populations of neural agents in the Learned Negation game, averaged over 10 repetitions, and excluding self-play. 95\% confidence intervals are shown in parentheses.}
    \label{avg-neural-learned-neg-no-self}   
\end{table}

\begin{table}[htb]
    \centering
    \begin{tabular}{|l|l|l|}
    \hline
        & $n=4$     & $n=8$     \\ \hline
    $p=2$ & 0.858 {\small (0.775, 0.941) \par} & 0.727 {\small (0.665, 0.810) \par} \\ \hline
    $p=5$ & 0.781 {\small (0.686, 0.876) \par} & 0.671 {\small (0.591, 0.752) \par} \\ \hline
    \end{tabular}
    \caption{Peak fitness for populations of Roth-Erev agents in the Combined Negation game, averaged over 10 repetitions, and excluding self-play. 95\% confidence intervals are shown in parentheses.}
    \label{avg-roth-erev-combined-neg-no-self}   
\end{table}

\begin{table}[tb]
    \centering
    \begin{tabular}{|l|l|l|}
    \hline
        & $n=4$     & $n=8$     \\ \hline
    $p=2$ & 0.803 {\small (0.721, 0.885) \par} & 0.670 {\small (0.585, 0.755) \par} \\ \hline
    $p=5$ & 0.639 {\small (0.611, 0.66) \par} & 0.570 {\small (0.498, 0.643) \par} \\ \hline
    \end{tabular}
    \caption{Peak fitness for populations of neural agents in the Combined Negation game, averaged over 10 repetitions, and excluding self-play. 95\% confidence intervals are shown in parentheses.}
    \label{avg-neural-combined-neg-no-self}   
\end{table}

As mentioned above, we do not observe substantial differences in performance between when self-play is included in evaluation and when it is excluded. In future work we hope to examine whether parameter sharing is helpful for the acquisition of compositional negation.

\end{document}